\documentclass[%
 aip,
 amsmath,amssymb,
 reprint,%
]{revtex4-1}

\usepackage{graphicx}
\usepackage{dcolumn}
\usepackage{bm}
\usepackage{xcolor}
\usepackage[utf8]{inputenc}
\usepackage[T1]{fontenc}
\usepackage{mathptmx}
\usepackage{etoolbox}
\usepackage[normalem]{ulem}
\usepackage{gensymb}

\makeatletter
\def\@email#1#2{
 \endgroup
 \patchcmd{\titleblock@produce}
  {\frontmatter@RRAPformat}
  {\frontmatter@RRAPformat{\produce@RRAP{*#1\href{mailto:#2}{#2}}}\frontmatter@RRAPformat}
  {}{}
}%
\makeatother

\begin{document}


\title[Visualizing metal-mediated nucleation and growth of GaN]{Visualizing metal-mediated nucleation and growth of GaN} 

\author{A. Liu}
\affiliation{Department of Materials Science and Engineering, University of Michigan, Ann Arbor, MI 48109, USA}
\author{Z. Xi}
\affiliation{Department of Materials Science and Engineering, University of Michigan, Ann Arbor, MI 48109, USA}
\author{X. Chen}
\affiliation{Department of Chemical and Petroleum Engineering, University of Pittsburgh, Pittsburgh, PA, USA}
\author{C. Huang}
\affiliation{Department of Materials Science and Engineering, University of Michigan, Ann Arbor, MI 48109, USA}
\author{M. Li}
\affiliation{Center for Functional Nanomaterials, Brookhaven National Laboratory, Upton, New York 11973, USA}
\author{J. C. Yang}
\affiliation{Department of Chemical and Petroleum Engineering, University of Pittsburgh, Pittsburgh, PA, USA}
\affiliation{Center for Functional Nanomaterials, Brookhaven National Laboratory, Upton, New York 11973, USA}
\affiliation{Department of Physics and Astronomy, University of Pittsburgh, Pittsburgh, PA, USA}
\author{L. Qi}
\affiliation{Department of Materials Science and Engineering, University of Michigan, Ann Arbor, MI 48109, USA}
\author{D. N. Zakharov}
\affiliation{Center for Functional Nanomaterials, Brookhaven National Laboratory, Upton, New York 11973, USA}
\author{R. S. Goldman}
 \thanks{Corresponding author: rsgold@umich.edu}
\affiliation{Department of Materials Science and Engineering, University of Michigan, Ann Arbor, MI 48109, USA}
\affiliation{Applied Physics Program, University of Michigan, Ann Arbor, MI 48109, USA}
\affiliation{Department of Physics, University of Michigan, Ann Arbor, MI 48109, USA}

\date{\today}

\begin{abstract}

Understanding the atomic-scale mechanisms governing metal-mediated nucleation and growth of gallium nitride (GaN) and related alloys is critical for tailoring their structural and functional properties in advanced electronic, optoelectronic, and quantum devices. Using real-time environmental transmission electron microscopy (E-TEM) in conjunction with Gibbs free energy calculations, we elucidate the distinct processes of 
GaN nucleation and growth
from Ga droplet arrays with and without GaN pre-nuclei.
For the lowest temperatures, although GaN nucleation at Ga droplet arrays is not observed, GaN growth occurs preferentially at pre-existing GaN nuclei, presumably due to the reduced Gibbs free energy for NH$_{3}$ decomposition at Ga/GaN interfaces. 
For intermediate  to high temperatures, 
E-TEM reveals nucleation and growth of GaN from Ga droplets with and without GaN nuclei, with enhanced crystallinity for the GaN nuclei, due to epitaxial templating.
These results highlight the critical role of the Ga/GaN interface in facilitating NH$_3$ decomposition and GaN growth, offering fundamental insights into metal-mediated nucleation and growth of GaN and related materials.


\end{abstract}


\maketitle 



\section{\label{sec:Intro}Introduction\protect\\}

Metal-mediated growth has played a central role in the synthesis of nanostructures, enabling control over morphology, composition, and polytype through liquid- or solid- phase mediation.\cite{shafi2022direct,heRecentAdvancesSeedMediated2023,niuSeedmediatedGrowthNoble2013,periwal2020catalytically} In these processes, metals act as catalysts, solvents, or structural templates that facilitate precursor decomposition, atomic diffusion, and crystallization. For example, liquid metal catalysts such as Au absorb vapor-phase precursors and precipitate solid material at the liquid–solid interface, leading to the formation of one-dimensional nanowires (NWs) through vapor-liquid-solid mechanisms.\cite{leeProducingAtomicallyAbrupt2017,ferrah2022photoemission} Self-catalyzed growth utilizing droplets such as Ga acts as catalysts to enable the formation of GaAs NWs through metal-mediated crystallization pathways.\cite{dubrovskii2021simultaneous,kang2025enhanced} In addition, droplet epitaxy, where pre-deposited metal droplets transform into crystalline nanostructures upon reaction with vapor-phase species, has been utilized to grow strain-free quantum dots in diverse semiconductor systems with tunable size and density.\cite{gurioliDropletEpitaxySemiconductor2019a,gajjela2022control}


It has been suggested that GaN growth occurs through metal-mediated mechanisms, where liquid Ga facilitates N incorporation and GaN nucleation and growth\cite{liu2024influence,lu2021influence,abdullah2023epitaxial}. While GaN has garnered significant interest due to its wide range of applications in high-power and high-frequency electronic devices \cite{heRecentAdvancesGaNBased2021,roblesRolePowerDevice2022a}, ultraviolet optoelectronics\cite{yao2023development,pu2023100}, and quantum technologies\cite{hoo2021emerging,de2002intrinsic}, 
only a few studies have employed in situ techniques to investigate GaN growth. For example, 
direct observations of early stages of Au-catalyzed GaN nanowire (NW) growth has been reported.\cite{diazDirectObservationNucleation2012}
Real-time observations of GaN NW growth via a multistep flow mechanism with double-bilayer steps, with Au-Ga particles serving as catalysts has also been reported.\cite{gamalskiAtomicResolutionSitu2016a} In another study, 
Self-catalyzed vapor-liquid-solid (VLS) growth following thermal decomposition of GaN has been examined.\cite{stachWatchingGaNNanowires2003} In this case, the Ga droplets and atomic N generated during the GaN thermal decomposition process enabled NW growth at pre-existing liquid Ga/solid GaN interfaces.\cite{stachWatchingGaNNanowires2003} To date, the distinct processes of nucleation and growth during self-catalyzed metal-mediated growth of GaN have not yet been visualized.


To distinguish nucleation and growth during NH$_{3}$-assisted metal-mediated growth of GaN, we examined the nucleation and growth process in real-time, with and without GaN pre-nuclei. Our findings suggest that the Ga/GaN interface plays a critical role in facilitating NH$_3$ decomposition by enhancing the driving force for nitrogen incorporation into Ga liquid, thereby accelerating GaN growth. In addition, we find a significant temperature dependence of the nucleation and growth processes. At the lowest temperatures, growth (without nucleation) is observed only at Ga/GaN interfaces, while at higher temperatures, GaN nucleation and growth are both observed within liquid Ga and at Ga/GaN interfaces, presumably due to enhanced NH$_{3}$ decomposition. These findings highlight the critical role of the Ga/GaN interface and the substrate temperature on decomposition of NH$_3$ and formation of GaN, providing fundamental insights into the mechanisms governing metal-mediated nucleation and growth of GaN and related materials.


\begin{figure*}
    \centering
    \includegraphics[width=\textwidth]{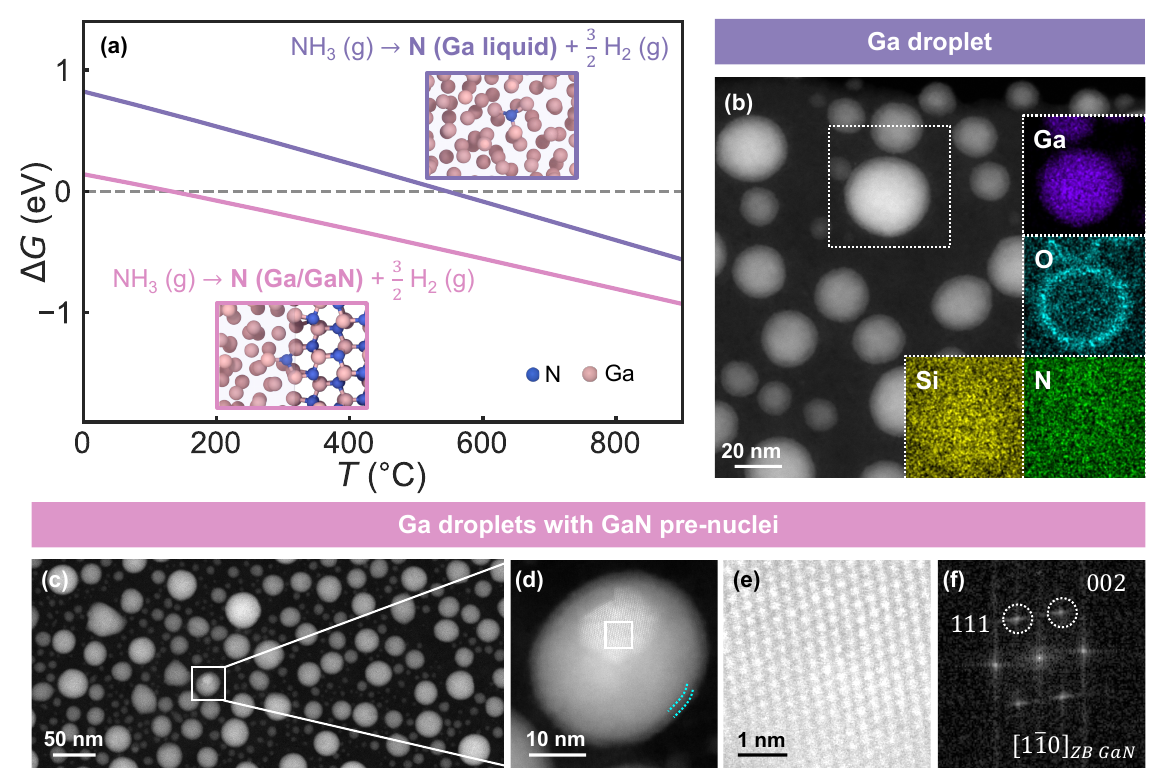}
    \caption{(a) Temperature-dependence of the Gibbs free energy, $\Delta G$, for NH$_3$ decomposition in liquid Ga (purple) and at the liquid Ga/GaN interface (pink). Intersections with the horizontal dashed line ($\Delta G = 0$) indicate the temperatures at which each pathway becomes thermodynamically favorable. For both pathways, the local environments used in the computation are shown in the insets: N adatoms in liquid Ga (top right) and at the GaN/Ga interface (bottom left).(b) Low-magnification high-angle annular dark field-scanning transmission electron micrograph (HAADF-STEM) of Ga droplet arrays, with  energy-dispersive spectroscopy (EDS) maps obtained from the white box region shown as insets. The EDS maps reveal droplets consisting of Ga-rich cores (purple) surrounded by thin oxide shells (blue), while the SiN substrate contains uniform concentrations of Si (yellow) and N (green). (c) Low-magnification HAADF-STEM of Ga droplet arrays with GaN pre-nuclei, with atomic-resolution HAADF-STEMs obtained from the white box regions in (d) and (f). The FFT from (f), shown in (g), is well-indexed to ZB GaN $[1\bar{1}0]$. }
\end{figure*}


\section{\label{sec:Methods}Materials and Methods\protect\\}
For these investigations, SiN chips (DENSsolutions) were etched in 5\% HF for 60 sec and mounted on specially-designed 3" sample blocks ("chips on blocks"). Following pre-baking for 8 hr in the load-lock and degassing for 30 min in the buffer chamber, the chips-on-blocks were transferred to the molecular-beam epitaxy chamber for deposition of Ga droplet arrays, with or without subsequent exposure to N plasma. Individual SiN chips containing Ga droplet arrays, with and without GaN pre-nuclei, were inserted into the differentially-pumped environmental transmission electron microscopy (E-TEM, FEI Titan 80–300) for real-time studies of GaN nucleation and growth at low (460~\degree\text{C}), intermediate (550~\degree\text{C}), and high (850~\degree\text{C}) temperatures. The source materials purities and fluxes are provided in Table S1 of the Supporting information. Additional post-growth scanning TEM (STEM) and energy-dispersive spectroscopy (EDS) were performed in a Thermo Fisher Scientific Talos F200X G2 S/TEM, equipped with a Super-X EDS detector. Identical imaging conditions were used for the Ga droplet arrays with and without GaN pre-nuclei.

To evaluate the driving force for NH$_3$ decomposition in liquid Ga and at the Ga/GaN interface, experimental thermodynamic data assessed in the CALculation of PHAse Diagrams (CALPHAD) framework~\cite{unland2003thermodynamics} and atomistic simulations based on the PreFerred Potential (PFP), a high-accuracy universal neural network interatomic potential \cite{takamoto2022universala,takamoto2023universala}, integrated within the Matlantis software package \cite{matlantis}, were utilized. The simulations were conducted using well-tempered metadynamics \cite{barducci2008well}, depositing tempered Gaussian bias along chosen collective variables to reconstruct the free-energy landscape, implemented via the PLUMED library,\cite{plumed2019promoting,tribello2014plumed} as described in the supplementary materials.

\begin{figure*}
    \centering
    \includegraphics[width=\textwidth]{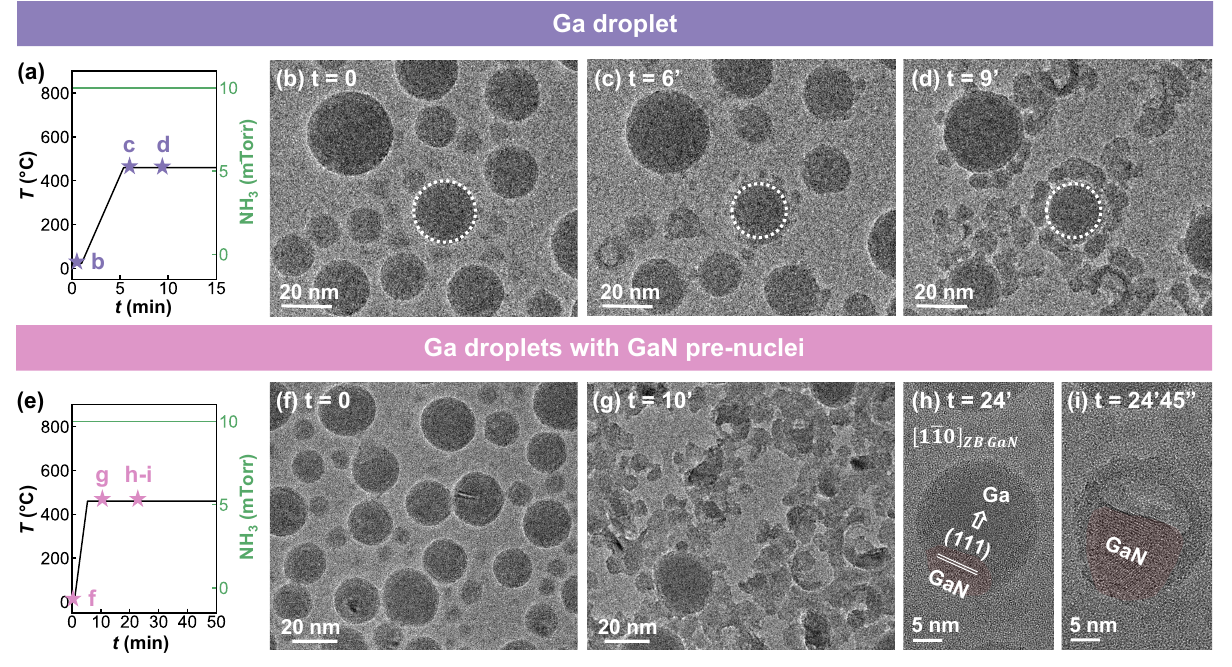}
    \caption{Nucleation and growth at the lowest temperatures: in-situ TEM images during NH$_3$ exposure with $\Delta T / \Delta t = 100~^{\circ}\mathrm{C}/\mathrm{min}$ and $T_{\text{target}}=460~\degree\text{C}$. (a) Heating and NH$_3$ exposure profiles for Ga droplet arrays. (b) Starting arrays of Ga droplets with oxide shells, with a tracking particle labeled with white dashed circles in (b)-(d). (c, d) Sequential TEM images at 6 min and 9 min showing the onset of surface restructuring and formation of small particles from Ga leakage following NH$_{3}$-etching of the Ga$_x$O$_y$ shells. (e) Heating and NH$_3$ exposure profiles for arrays of Ga droplet with GaN pre-nuclei. (f) Starting arrays of Ga droplets with GaN pre-nuclei. (g) Low-magnification TEM image of arrays of Ga droplets with GaN pre-nuclei after exposure of NH$_3$ at 10 min. Time-resolved TEM images at (h) 24 min and (i) 24 min 45 sec, showing progressive lateral GaN growth (colored in red) along the Ga/GaN interface.}
\end{figure*}


\section{\label{sec:Results}Results and Discussion\protect\\}

The temperature-dependence of the Gibbs free energy, $\Delta G$, for NH$_3$ decomposition in liquid Ga and at the liquid Ga/GaN interface are shown in Fig. 1(a).  For all temperatures, the $\Delta G$ values are lowest for the liquid Ga/GaN interfaces. In the plot of $\Delta G$ vs. T, the horizontal dashed line denotes $\Delta G = 0$, and its intersections mark the temperature at which each pathway becomes thermodynamically favorable ($\Delta G < 0$). At the liquid Ga/solid GaN interface (pink line), $\Delta G$ becomes negative at $\sim$ 150~\degree\text{C}, while in liquid Ga (purple line), $\Delta G$ only becomes negative at a intermediate temperature of 550~\degree\text{C}. Thus, a higher driving force is required for the NH$_3$ dissociation and subsequent growth of GaN in Ga liquid in comparison to that at the liquid Ga/solid GaN interface.

Examples of Ga droplet arrays for E-TEM studies are shown in the low-magnification HAADF-STEM images in Figs. 1 (b) - (f).
For Ga droplet arrays (without pre-nuclei), shown in Fig. 1(b), circular Ga droplets surrounded by thin ($\sim$2 nm), low-contrast layers  are apparent. For the white box region in Fig. 1(b), EDS maps (in the insets) reveal droplets consisting of Ga-rich cores (purple) surrounded by a thin oxide shells (blue), while the SiN substrate contains uniform concentrations of Si (yellow) and N (green). For the Ga droplet arrays with GaN pre-nuclei, shown in Fig. 1(c), the circular morphology of the Ga droplets is retained, with a fraction of each Ga droplet converted to GaN. In this case, an oxide shell, highlighted by dotted lines in Fig. 1(d), is also observed external to the Ga/GaN particles. Figs. 1(e) and 1(f) show the atomic structure and the FFT for Ga droplets with GaN pre-nuclei, which together confirm the presence of GaN.  Although the FFTs for this example are indexed to zinc blende (ZB) GaN $[1\bar{1}0]$, the wurtzite (WZ) polytype is also frequently observed.

\begin{figure*}
    \centering
    \includegraphics[width=\textwidth]{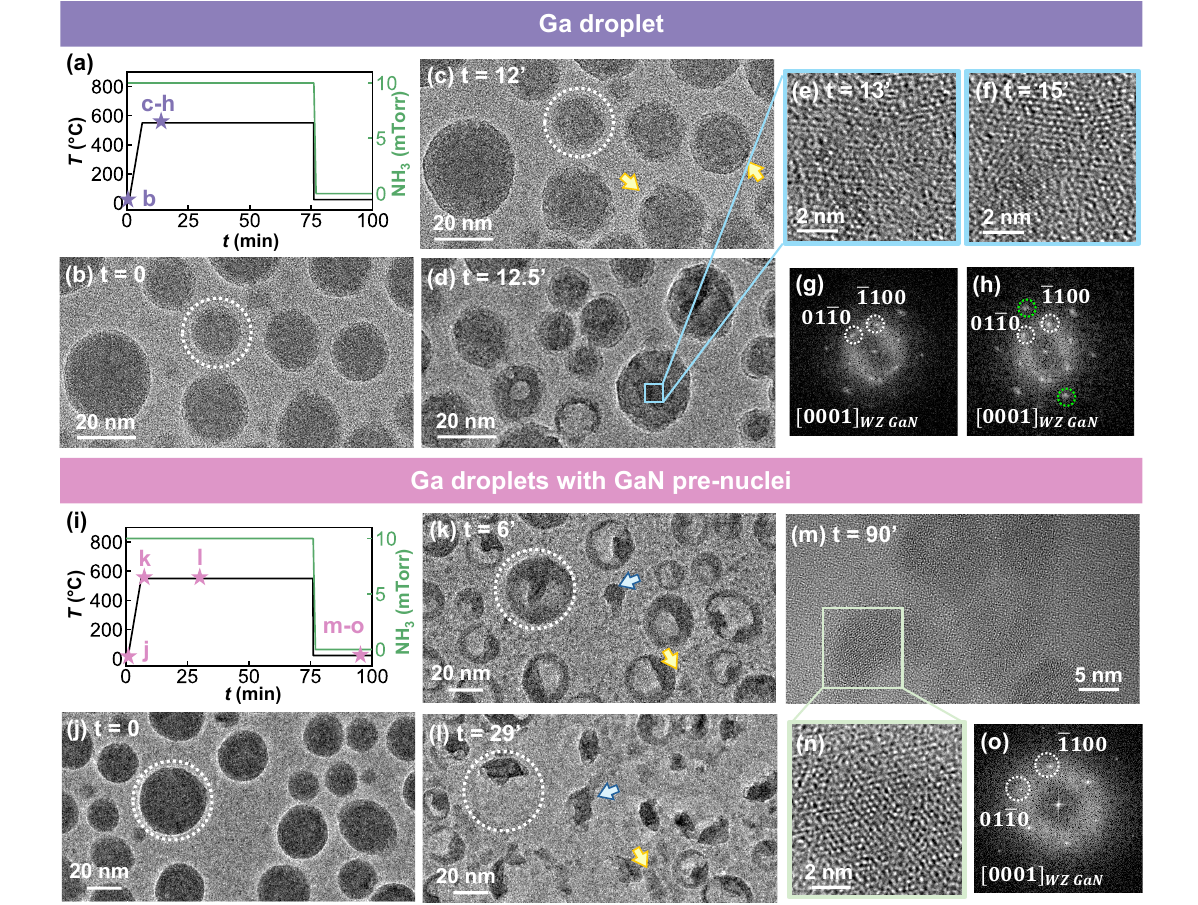}
    \caption{Nucleation and growth at intermediate temperature: in-situ TEM images during NH$_3$ exposure with $\Delta T / \Delta t = 100~^{\circ}\mathrm{C}/\mathrm{min}$ and $T_{\text{target}}=550~\degree\text{C}$. (a) Heating and NH$_3$ exposure profiles for Ga droplet arrays. Real-time TEM images (b) at t = 0 and (b) at 12 min. The yellow arrows  in (c) mark the darker contrast on the surface. The surfaces of Ga droplet arrays have roughened, likely due to the etching of the oxide shell by NH$_3$. (d) Darker contrast appears at the center of the particle, with atomic resolution images of the blue-boxed region showing in (e) 13 min and (f) 15 min. The FFTs taken from (e) and (f) are both indexed to (g, h) WZ GaN $[0001]$. (i) Heating and NH$_3$ exposure profiles for arrays of Ga droplets with GaN pre-nuclei. Real-time TEM images of the arrays of Ga droplets with GaN pre-nuclei (j) at t = 0, (k) at 6 min and (l) at 29 min. Ga begins to desorb from smaller Ga/GaN particles, leaving oxide shells on the substrate. Reduced contrast between the Ga/GaN particles and the SiN substrate is likely due to Ga wetting the substrate surface. Tracked particles are marked in each frame with white dotted circles, blue arrows, and yellow arrows for reference. (m) HRTEM image of a GaN particle formed in E-TEM. (n) provides a close-up view, with the FFTs of (n) indexed to (o) WZ GaN $[0001]$ orientation.}
\end{figure*}

We now present nucleation and growth of GaN from Ga droplet arrays at low temperature (460~\degree\text{C}), with the heating and NH$_3$ exposure profiles shown in Fig. 2(a), and a tracking particle labeled with white dashed circles in Fig. 2(b)-(d). The as-prepared arrays of Ga droplets (Fig. 2(b)) exhibit a circular morphology with shells consisting of a uniform amorphous oxide layer, consistent with the features shown in Fig. 1(b). Upon NH$_3$ introduction and heating (Figs. 2(c)-(d)), the droplet surfaces are restructured, with 
ultra-small "satellite" particles (< 5 nm in diameter) appearing on the droplet perimeters, presumably due Ga leakage following NH$_{3}$-etching of the Ga$_x$O$_y$ shells.  Sequential TEM images at 
6 min (Fig. 2(c)) and  9 min (Fig. 2(d)) reveal increasing sizes of the "satellite" particles. 
This phenomenon is further supported by the low-magnification overview in the Supporting information Fig. S2(b), which shows the nucleation of small particles on the perimeter of Ga droplet arrays. A high-resolution TEM image (Fig. S2(c)) highlights these surface particles, revealing amorphous nanoparticles on the droplet surface without evidence of crystalline GaN formation. The corresponding FFT pattern (Fig. S2(d)) 
displays diffuse features characteristic of disordering.

We next consider the nucleation and growth from the arrays of Ga droplets with GaN pre-nuclei at low temperature, with the heating and NH$_3$ exposure profiles shown in Fig. 2(e). At $t = 0$ (Fig. 2(f)), due to the presence of GaN pre-nuclei, darker regions are observed within Ga droplet arrays. At $t = 10$ min, Ga leakage and small particles are observed, similar to Fig. 2(d). In contrast to the low temperature nucleation and growth from Ga droplet arrays, directional GaN growth is apparent. For example, at $t=24$ min, Fig. 2(h) illustrates an HRTEM image of a Ga droplet with GaN pre-nuclei (colored in red), viewed along the $[1\bar{1}0]$ zone axis with lattice fringes of GaN indexed to be the (111) planes of ZB GaN. Subsequent time-resolved HRTEM imaging (Figs. 2(i)) reveals progressive GaN thickening via lateral propagation along the Ga/GaN interface, demonstrating NH$_3$ dissociation enabled by the liquid Ga/GaN interfaces, with 
the nucleation of new atomic steps at the interface and their lateral advancements shown in Fig. S2.

\begin{figure*}
    \centering
    \includegraphics[width=\textwidth]{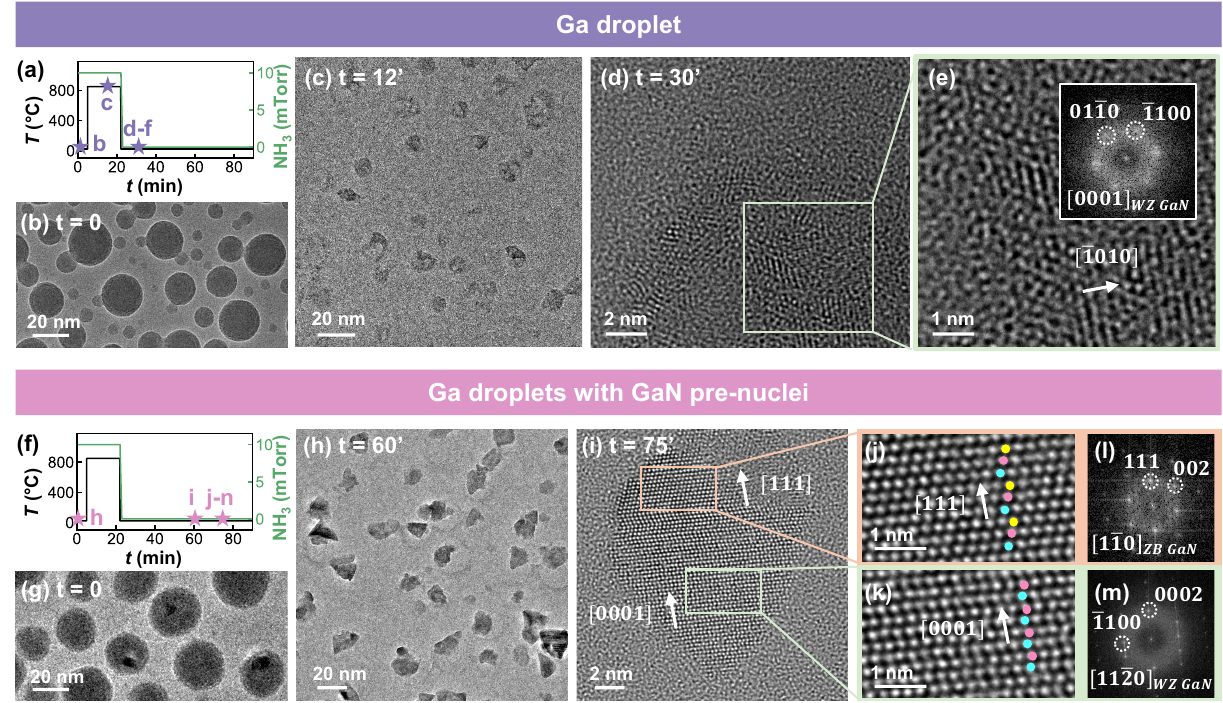}
    \caption{Nucleation and growth at high temperature: in-situ TEM images during NH$_3$ exposure with $\Delta T / \Delta t = 400~^{\circ}\mathrm{C}/\mathrm{sec}$ and $T_{\text{target}}=850~\degree\text{C}$. (a) Heating and NH$_3$ exposure profiles for Ga droplet arrays. (b-c) The evolution of Ga droplet arrays from circular to irregular shape during exposure to NH$_3$. (d) HRTEM image of an example particle showing WZ GaN formation. (e) provides a close-up view, with an FFT shown in the inset, further confirming the formation of WZ GaN. (f) Heating and NH$_3$ exposure profiles for arrays of Ga droplet with GaN pre-nuclei. The arrays of Ga droplets with GaN pre-nuclei transform from (g) circular morphology prior to growth into (h) triangular shape following growth. (i) Atomic-resolution HRTEM image of  GaN reveals a mixed polytype GaN nanostructure that transitions from the ZB to the WZ polytypes. (j) Atomic-resolution HRTEM image from the orange-boxed regions in (i), with blue, pink, and yellow dots illustrating ABC stacking along the $[111]$ direction. (k) Atomic-resolution image from the green-boxed regions in (i), with blue and pink dots representing ABAB stacking along the [0001] direction. The FFTs taken from orange- and green-boxed regions are indexed to (l) ZB GaN $[1\bar{1}0]$ and (m) WZ GaN $[11\bar{2}0]$, respectively, with ZB $[111]$ parallel to WZ $[0001]$.}
\end{figure*}


We now present nucleation and growth of GaN from Ga droplet arrays at intermediate temperature (550~\degree\text{C}), with the heating and NH$_3$ exposure profiles shown in Fig. 3(a).  
Figures 3(b-f) show real-time TEM images capturing GaN growth from Ga droplet arrays without GaN pre-nuclei.  
At $t = 0$ (Fig. 3(b)), Ga droplet arrays exhibit a circular shape with uniform contrast across all droplets, similar to those in Fig. 1(b). At $t = 12$ min (Fig. 3(c)), the surfaces of the Ga droplet arrays have roughened, likely due to the etching of the oxide shell by NH$_3$. Additionally, darker contrast regions on the surface of several droplets, indicated by yellow arrows, may indicate the formation of GaN nucleation sites.

\begin{figure*}
    \centering
    \includegraphics[width=\textwidth]{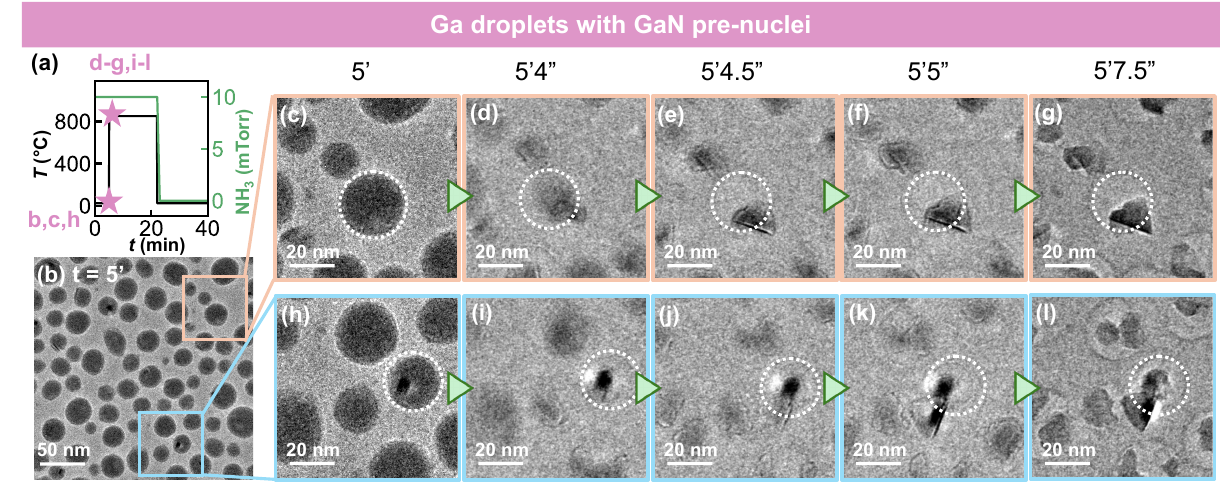}
    \caption{In situ TEM images during NH$_3$ exposure with $\Delta T / \Delta t = 400~^{\circ}\mathrm{C}/\mathrm{sec}$ and $T_{\text{target}}=850~\degree\text{C}$. (a) Heating and NH$_3$ exposure profiles for arrays of Ga droplets with GaN pre-nuclei. (b) Low-magnification TEM image of the arrays of Ga droplets with GaN pre-nuclei. Real-time TEM images from 5 min to 5 min 7.5 sec (c-g) showing faceted particles that grow along GaN pre-nuclei and (h-l) nucleation and growth of GaN. White dotted circles mark the same positions for reference. Slight focus shift occurs due to rapid heating.}
\end{figure*}

In a nearby region, at $t = 12.5$ min (Fig. 3(d)), Ga begins to desorb from smaller droplets, leaving behind the oxide shell on the substrate. 
Interestingly, in Fig. 3(d), a particle with darker contrast at its center is apparent; within the region highlighted by the blue box, atomic-resolution images and corresponding FFTs 
are shown at 13 min (Figs. 3(e), (g)) and 15 min (Figs. 3(f), (h)) 
At $t=13 $ min, lattice fringes emerge (Fig. 3(e)), and the FFTs 
reveal the formation of wurtzite (WZ) GaN along $[0001]$ zone axis (Fig. 3(g)).  As the reaction progresses to 15 min, the crystallinity of the WZ GaN improves through continued atomic rearrangement, as evidenced by the HRTEM image (Fig. 3(f)) and corresponding FFT pattern (Fig. 3 (h)). The appearance of two additional reflection spots, marked by green dotted circles, corresponds to higher-order WZ reflections, indicative of improved crystallinity.

For comparison, we consider the 
nucleation and growth at intermediate temperature for the arrays of Ga droplets with GaN pre-nuclei, using the heating and NH$_3$ exposure profiles in Figs. 3(i). At $t=0$, due to the presence of GaN pre-nuclei, the contrast within the particles is non-uniform, with GaN appearing darker in the TEM image (Fig. 3(j)). In Figs 3(j-l), tracked particles are marked in each frame with white dotted circles, blue arrows, and yellow arrows for reference. 

As the temperature increases from room temperature (Fig. 3(j)) to 550~\degree\text{C} (Fig. 3(k)), Ga begins to desorb from smaller Ga/GaN particles, leaving oxide shells on the substrate, similar to the case of Ga droplet arrays (Fig. 3(d)). By $t=6$ min (Fig. 3(k)), Ga has desorbed from most Ga/GaN particles, with reduced contrast between the Ga/GaN particles and the SiN substrate, likely due to Ga wetting the substrate surface. Additionally, two new particles, marked by blue and yellow arrows in Fig. 3(k), have appeared, presumably due to the leakage and migration of Ga from the Ga/GaN particles.
For the particle marked by the blue arrow, a transformation from circular to faceted shape is apparent (Fig. 3(k) to Fig. 3(l)), suggesting that crystallization has occurred. Meanwhile, for the particle labeled by the yellow arrow, an increase in size from Fig. 3(k) to Fig. 3(l) is apparent, presumably due to liquid Ga aggregation followed by crystallization. 

On the other hand, for the Ga/GaN particle marked by a white dotted line (Fig. 3(k)), 
the upper region becomes faceted following Ga desorption (Fig. 3(l)). Fig. 3(m) provides an example of a GaN particle 
following cooling to 20~\degree\text{C}, 
with a zoom-in view of the green box region displayed in Fig. 3(n). The FFTs of the green box region (Fig. 3(o)) are indexed to WZ GaN $[0001]$. Indeed, increasing the temperature from 460~\degree\text{C} to 550~\degree\text{C} leads to GaN nucleation and growth, as evidenced in Figs. 3(e) and 3(n). Thus, for intermediate temperature, there is adequate NH$_3$ decomposition for GaN growth in both liquid Ga and at Ga/GaN interfaces. 

We now consider nucleation and growth of GaN from Ga droplets at high temperature, with the heating and NH$_3$ exposure profiles shown in Fig. 4(a).
Figs. 4(b, c) illustrate the evolution of Ga droplet arrays from circular to irregular shape during exposure to NH$_3$ at 850 \degree\text{C}. For an example particle shown in the HRTEM image (Fig. 4(d)), WZ GaN formation, with limited crystallinity, is apparent.  The close-up view of the green box region in Fig. 4(d) is shown in Fig. 4(e), with an FFT shown in the inset.  The close-up view (Fig. 4(e))
further confirms the formation of WZ GaN, while the intermittent disappearance of atomic columns in the FFT inset suggests the involvement of defects during crystal growth.  

For the arrays of Ga droplets with GaN pre-nuclei, the heating and NH$_3$ profiles for nucleation and growth at high temperature are shown in Fig. 4(g). For this case, the Ga droplet arrays with GaN pre-nuclei transform from circular morphologies (Fig. 4(h)) into triangular shapes (Fig. 4(i)). Further post-growth details are provided by the atomic-resolution image in Fig. 4(j), which reveals a mixed polytype GaN nanoparticle that transitions from the ZB to the WZ polytypes. For the orange-boxed region in Fig. 4(j), the corresponding HRTEM image in Fig. 4(k) is overlayed with blue, pink, and yellow dots illustrating ABC stacking along $[111]$ direction, suggesting the presence of the ZB polytype.  The presence of the ZB polytype is further supported by the corresponding FFT (Fig. 4(l)) which is indexed to the ZB GaN $[1\bar{1}0]$. For the green-boxed region in Fig. 4(j), the corresponding HRTEM image in Fig. 4(m) is overlayed with blue and pink dots illustrating ABAB stacking along $[0001]$ direction, characteristic of the WZ polytype. The presence of the WZ polytype is further supported by the corresponding FFT (Fig. 4(n)) which is indexed to the WZ GaN $[0001]$.  
Additional HRTEM images of GaN nanoparticles prepared at high temperatures are presented in Fig. S4. Compared to the pure Ga droplet arrays, the arrays of Ga droplets with GaN pre-nuclei display significantly improved crystallinity and more regular morphologies, consistent with kinetically accelerated GaN growth process. 

For further insight into the nucleation and growth processes at high temperatures, we consider snapshots from real-time TEM images collected along the arrays of Ga droplets with GaN pre-nuclei (Fig. 5), with the heating and NH$_3$ exposure profiles shown in Fig. 5(a). From the low-magnification TEM image shown in Fig. 5(b), we track the evolution of several Ga/GaN particles within the orange-boxed (Fig. 5(c-g)) and blue-boxed (Fig. 5(h-l)) regions. White dotted circles in Figs. 5(c–l) indicate the same reference positions throughout the sequence. At 5 min, the dark circular regions correspond to Ga droplet arrays, with darker contrast regions indicating GaN pre-nuclei. Due to the rapid heating (the temperature increases to 850~\degree\text{C} within ~2 s), a slight focus shift is apparent. 
For the orange-boxed region, between 5 min and 5 min 4 s, Ga desorbs from the droplets, consistent with the observations at lower temperatures (Fig. 3(k)). Subsequently, GaN growth proceeds along the GaN pre-nuclei sites, as shown in Figs. 5(d–g). Along the droplet edge, crystalline contrast emerges, with the formation of a triangular GaN particle shown in Fig. 5(g). The extension of the triangular region beyond the white dotted circle in Fig. 5(g) confirms that it is not solely GaN pre-nuclei but rather newly crystallized GaN grown epitaxially along the existing Ga/GaN interface under NH$_3$ exposure.

In addition, for the blue-boxed region, Figs. 5(h–l) show independent GaN nucleation near a Ga droplet. In Fig. 5(h), non-uniform contrast is apparent within the droplet, where the darker region corresponds to GaN pre-nuclei and the brighter region to liquid Ga. Upon heating, partial Ga desorption occurs (Figs. 5(i-j)), and a new crystalline feature emerges along the lower edge of the GaN pre-nuclei (Fig. 5(k)). As growth proceeds, the newly nucleated GaN adopts a triangular morphology, whereas the nearby GaN pre-nuclei evolves into hexagon-like morphology. Eventually, the two crystallites coalesce, forming a larger crystalline structure (Fig. 5(l)). This behavior demonstrates that GaN formation can occur simultaneously through epitaxial growth of existing GaN pre-nuclei and through independent nucleation and growth.
 


We now discuss the influence of the temperature and Ga/GaN interface on the GaN growth. At low temperature, NH$_3$ decomposition is thermodynamically favorable only at Ga/GaN interface, as shown in Fig. 1(a). The absence of atomic N in liquid Ga leads to no GaN nucleation and growth occurs in Ga droplet arrays, while the presence of atomic N at Ga/GaN interface results in GaN growth, as confirmed in Fig. 2. On the other hand, at intermediate temperature, NH$_3$ decomposition becomes favorable at both liquid Ga and the Ga/GaN interface,  providing a sufficient supply of atomic N to drive GaN formation, as evidenced in Fig. 3.  At high temperature, faceting of the GaN particles becomes more pronounced (Fig. 4(i)), highlighting the critical role of temperature in promoting crystalline growth.

Interestingly, although GaN growth occurs in the arrays of Ga droplets with and without GaN pre-nuclei at intermediate temperature, the resulting particle morphologies are different, as shown in Fig S3. In the Ga droplet arrays (Fig. S3(b)), large aggregated particles are observed, whereas such aggregates are absent in the arrays of Ga droplets with GaN pre-nuclei (Fig. S3(j)). HRTEM analysis (Fig. S3(c-h)) further reveals the coexistence of GaN and Ga$_{x}$O$_{y}$ within the aggregated particles in Fig. S3(b). We hypothesize that at intermediate temperature, once NH$_3$ etches the oxide shell, Ga does not immediately desorb; instead, it either nucleates into GaN within the shell or migrates across the substrate, where it subsequently aggregates and nucleates into GaN or Ga$_{x}$O$_{y}$. In contrast, in the arrays of Ga droplets with GaN pre-nuclei, the presence of the Ga/GaN interface facilitates NH$_3$ decomposition, thereby generating a higher concentration of active nitrogen species at the interface. This enhanced NH$_3$ dissociation promotes GaN crystallization while simultaneously suppressing Ga migration and aggregation.

\section{\label{sec:Sum}Summary\protect\\}

In summary, combining free energy calculations with real-time E-TEM observations, we reveal the critical role of the Ga/GaN interface in governing NH$_3$ decomposition and subsequent GaN growth. Free energy calculations demonstrate that the Ga/GaN interface lowers the onset temperature for NH$_3$ decomposition, which in turn enhances N incorporation into liquid Ga as needed for GaN growth.
Furthermore, real-time E-TEM observations reveal that at low temperature, GaN growth occurs only at Ga/GaN interfaces, while Ga droplet arrays show only broken oxide shell and Ga leakage without crystallized into GaN. At intermediate temperatures, NH$_3$ decompose more easily in Ga liquid, resulting in nucleation and growth that occurs in the arrays of Ga droplets with and without GaN pre-nuclei. In addition, at high temperature, more faceted GaN particles with better crystallinity are observed using the arrays of Ga droplets with GaN pre-nuclei. These results highlight the important role of Ga/GaN interface and temperature to NH$_3$ decomposition and GaN nucleation and growth, elucidating key mechanisms underlying metal-mediated growth of GaN and related systems.

\begin{acknowledgments}
This work was supported by the Department of Energy, Basic Energy Sciences under Award No. DE-SC0023222. The authors acknowledge the assistance of the staff at the Michigan Center for Materials Characterization at the University of Michigan. This research used Electron Microscopy resources of the Center for Functional Nanomaterials (CFN), which is a U.S. Department of Energy Office of Science User Facility, at Brookhaven National Laboratory under Contract No. DE-SC0012704.
\end{acknowledgments}


\section*{Data Availability Statement}
Data available in article or supplementary material.

\section*{References}
\bibliography{References}


%
%

%


\end{document}



\title[Visualizing metal-mediated nucleation and growth of GaN]{Supplementary Material\\Visualizing metal-mediated nucleation and growth of GaN} 



\author{Abby Liu}
\affiliation{Department of Materials Science and Engineering, University of Michigan, Ann Arbor, MI 48109, USA}
\author{Zhucong Xi}
\affiliation{Department of Materials Science and Engineering, University of Michigan, Ann Arbor, MI 48109, USA}
\author{Xiaobo Chen}
\affiliation{Center for Functional Nanomaterials, Brookhaven National Laboratory, Upton, New York 11973, USA}
\author{Catherine Huang}
\affiliation{Department of Materials Science and Engineering, University of Michigan, Ann Arbor, MI 48109, USA}
\author{Meng Li}
\affiliation{Center for Functional Nanomaterials, Brookhaven National Laboratory, Upton, New York 11973, USA}
\author{Judith C. Yang}
\affiliation{Center for Functional Nanomaterials, Brookhaven National Laboratory, Upton, New York 11973, USA}
\affiliation{Department of Chemical and Petroleum Engineering, University of Pittsburgh, Pittsburgh, PA, USA}
\affiliation{Department of Physics and Astronomy, University of Pittsburgh, Pittsburgh, PA, USA}
\author{Liang Qi}
\affiliation{Department of Materials Science and Engineering, University of Michigan, Ann Arbor, MI 48109, USA}
\author{Dmitri N. Zakharov}
\affiliation{Center for Functional Nanomaterials, Brookhaven National Laboratory, Upton, New York 11973, USA}
\author{Rachel S. Goldman}
 \thanks{Corresponding author: rsgold@umich.edu}
\affiliation{Department of Materials Science and Engineering, University of Michigan, Ann Arbor, MI 48109, USA}
\affiliation{Applied Physics Program, University of Michigan, Ann Arbor, MI 48109, USA}
\affiliation{Department of Physics, University of Michigan, Ann Arbor, MI 48109, USA}

\date{\today}

\begin{abstract}
We present details of the free energy calculations for the ammonia decomposition and nitrogen adsorption processes. In addition, the purity of the source materials, as well as the fluxes and exposure times during molecular-beam epitaxy (MBE) and environmental transmission electron microscopy (E-TEM) are provided. Finally, supplemental transmission electron microscopy images from real-time studies of GaN nucleation/growth at low, intermediate, and high temperatures are presented.
\end{abstract}

\pacs{}

\maketitle 
\renewcommand\thesection{S\Roman{section}}

\section{Ammonia Decomposition Calculations \protect\\}
Here, we present the thermodynamic calculations for NH$_3$ decomposition in liquid Ga and at the Ga/GaN interface using the temperatures and pressures of the E-TEM experiments. We first consider the free energy for NH$_3$ decomposition in the gas phase 

\begin{equation}
\label{eq:re1}
    \text{NH}_3 (\text{g}) \to \frac{3}{2}\text{H}_2 (\text{g}) + \frac{1}{2}\text{N}_2 (\text{g}).
\end{equation}

\begin{figure}[!h]
    \centering
    \includegraphics[width=1\linewidth]{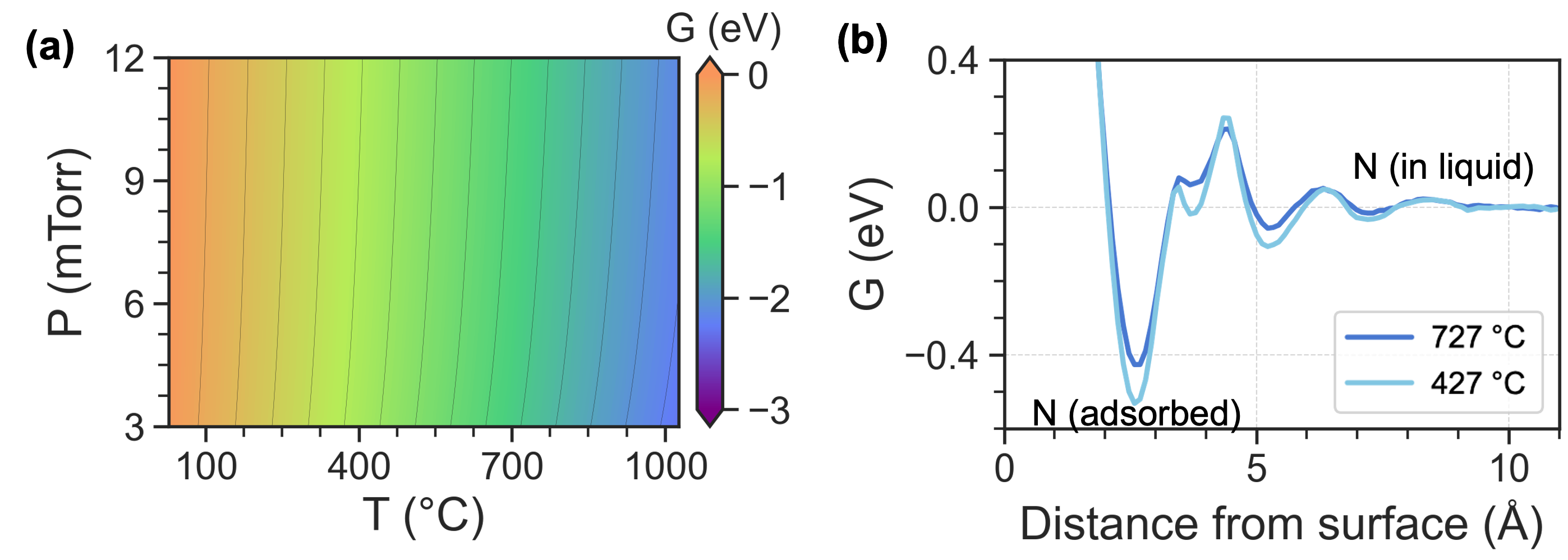}
    \caption{(a) Free energy contour for ammonia dissociation at various temperatures and pressures. (b) Free energy profiles obtained from metadynamics for the reaction \ref{eq:re4} (d) at 427~\degree\text{C} and 727~\degree\text{C}.}
    \label{fig:FreeEnergy}
\end{figure}

Assuming similar partial pressures for NH$_3$, N$_2$, and H$_2$, we use the ideal gas model to compute the free energy for NH$_3$ dissociation as a function of temperature and pressure. The resulting plot of free energy as a function of pressure and temperature is shown in Fig. \ref{fig:FreeEnergy}(a), where free energies ranging from -3 to 0 eV are shown as colors ranging from purple to orange. Indeed, the NH$_{3}$ dissociation reaction is most favorable at high-temperature and low-pressures, where the free energies are negative. 

Next, we consider the free energy for N$_2$ solvation 


\begin{equation}
\label{eq:re2}
    \text{N}_2 (\text{g}) \to 2 \text{N} (\text{Ga liquid}).
\end{equation}
In this case, the Ga–rich liquid is treated as a binary regular solution, and the solvation energy of N in the liquid is parameterized as $G_{\mathrm{N}}^{\mathrm{Ga\,liquid}}(T)-\frac{1}{2}G_{\mathrm{N_2}}^{\mathrm{gas}}(T) = 0.656 + 6.117\times10^{-4} \,T$ eV/atom.\cite{unland2003thermodynamics}
Finally, metadynamics simulations are used to compute the free energy for N adsorption resulting in GaN growth 
\begin{equation}
\label{eq:re4}
\text{N} (\text{Ga liquid}) \to \text{N} (\text{adsorbed to GaN}).
\end{equation}
For the metadynamics simulations, the Ga/GaN ($000\bar{1}$) N-terminated interface was constructed using a solid GaN slab with 384 Ga and 384 N atoms, and a Ga liquid region with 500 Ga atoms. A single N atom was placed at the GaN/Ga interface, with a one-dimensional collective variable defined as its distance to the solid GaN surface. The system was equilibrated for 10 ps at 1000 K and at 700K using a Langevin thermostat.\cite{schneider1978molecular} During each 10 ns simulation, Gaussians (0.1 Å width, 0.01 eV height) were deposited every 0.5 ps with a bias factor of 5. 

Fig. \ref{fig:FreeEnergy}(b) presents the free energy change for a single adatom N atom approaching the GaN/Ga interface along the $\langle000\bar{1}\rangle$ N-polar interfaces at 427~\degree\text{C} and 727~\degree\text{C}. Both profiles exhibit deep energy minima ($\sim$-0.5 to -0.4 eV), implying stronger binding and more efficient decomposition of ammonia at the interfaces. Temperature-dependent free energies for binding are determined from a linear fit to these data points.

\section{Source Materials and Fluxes}

\begin{table}[h]
    \centering
    \caption{The purity of the source materials, fluxes, and exposure times during MBE and E-TEM.}
    \renewcommand{\arraystretch}{1.2}
    \setlength{\tabcolsep}{4pt} 
        \begin{tabular}{|l|l|l|l|l|l|}
            \hline
            \textbf{Process} & \textbf{Source} & \textbf{Purity (\%)} & \textbf{Flux (Torr)} & \textbf{Time} & \textbf{Temp (°C)} \\ \hline
            Ga droplets       & Ga     & 99.99999 & $4.4 \times 10^{-7}$ & 53 s      & 650 \\ \hline
            MBE nitridation   & N$_2$  & 99.9999  & $5.5 \times 10^{-7}$ & 60 s      & 650 \\ \hline
            E-TEM nitridation & NH$_3$ & 99.99    & $10^{-2}$            & > 20 min  & 460, 550, 850 \\ \hline
        \end{tabular}

    \label{tab:placeholder_label}
\end{table}

\section{Supplementary E-TEM Data\protect\\}

In this section, we present supplementary transmission electron microscopy data for GaN nucleation and growth at low (Figure S2), intermediate (Figure S3), and high (Figure S4) temperatures.

\begin{figure}
    \centering
    \includegraphics[scale=0.95]{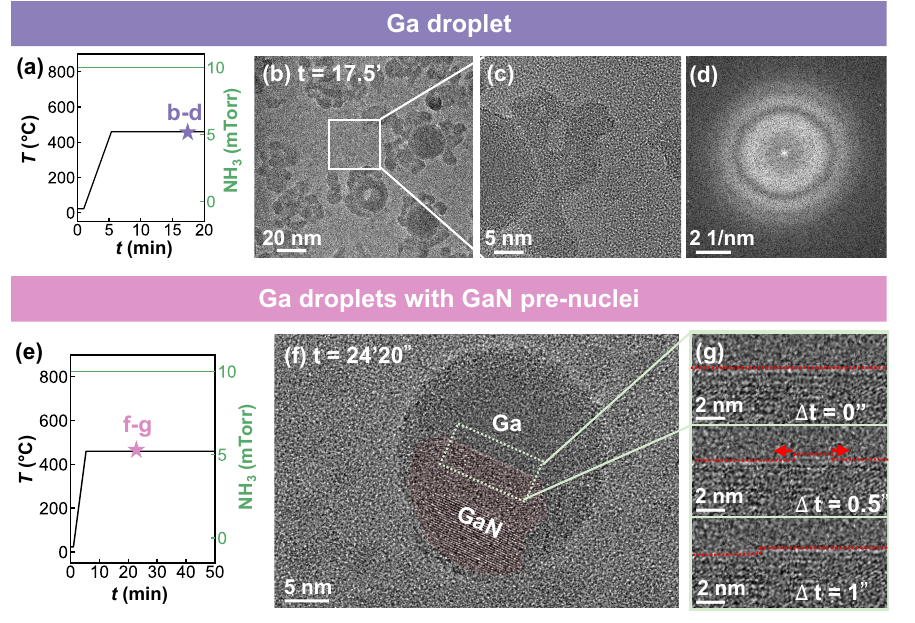}
    \caption{\label{fig:S1-ETEM} Nucleation and growth at low temperature: \textit{supplementary} in-situ TEM images during NH$_3$ exposure with $\Delta T / \Delta t = 100~^{\circ}\mathrm{C}/\mathrm{sec}$ and $T_{\text{target}}=460~\degree\text{C}$ (a) Heating and NH$_3$ exposure profiles for Ga droplet arrays. (b) Low-magnification TEM image reveals nucleation of small particles on the perimeter of Ga droplets. (c) Close-up view from white box region in (b), showing surface nanoparticles. (d) Corresponding FFT of (c) with diffuse features characteristic of disordering. (e) Heating and NH$_3$ exposure profiles for arrays for Ga droplets with GaN pre-nuclei. (f) An example of GaN growth with (g) subsequent time-resolved HRTEM imaging showing progressive GaN thickening via lateral propagation along the Ga/GaN interface.}
\end{figure}

\begin{figure}
    \centering
    \includegraphics{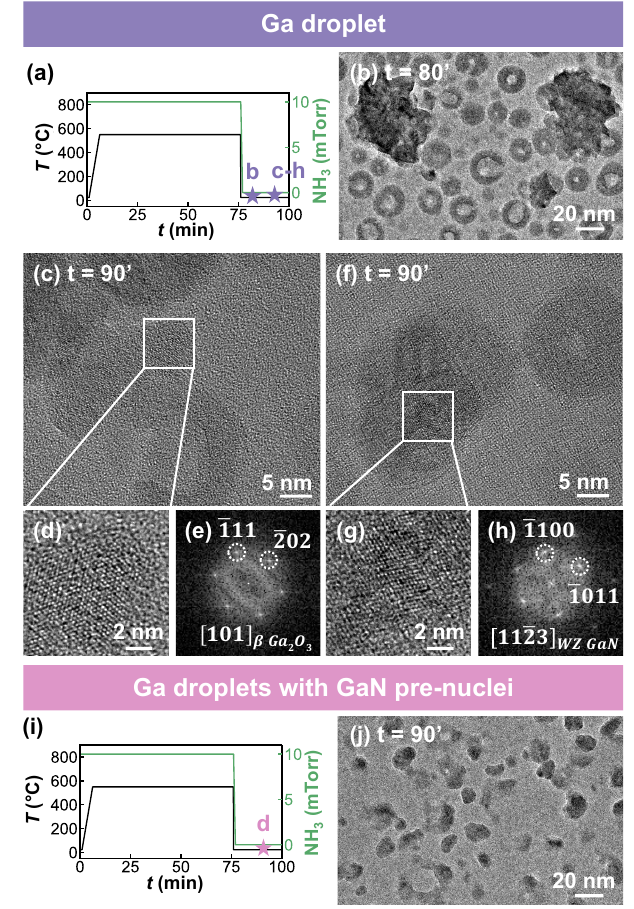}
    \caption{\label{fig:S3-ETEM} Nucleation and growth at intermediate temperature: \textit{supplementary} in-situ TEM images during NH$_3$ exposure with $\Delta T / \Delta t = 100~^{\circ}\mathrm{C}/\mathrm{sec}$ and $T_{\text{target}}=550~\degree\text{C}$ (a) Heating and NH$_3$ exposure profiles of Ga droplets. (b) Low-magnification TEM images after NH$_3$ exposure of Ga droplets, revealing the formation of large-aggregated particles. (c, f) HRTEM images of particles after NH$_3$ exposure. (d) and (g) provide close-up views, and the FFTs taken from (d) and (g) are indexed to (e) $\beta$-Ga$_{2}$O$_{3}$ $[101]$ and (h) WZ GaN $[11\bar{2}3]$, respectively. (i) Heating and NH$_3$ exposure profiles of pre-nucleated Ga/GaN particles. (j) Low-magnification TEM images after NH$_3$ exposure of pre-nucleated Ga/GaN.}
\end{figure}

\begin{figure}[!h]
    \centering
    \includegraphics[scale=1.05]{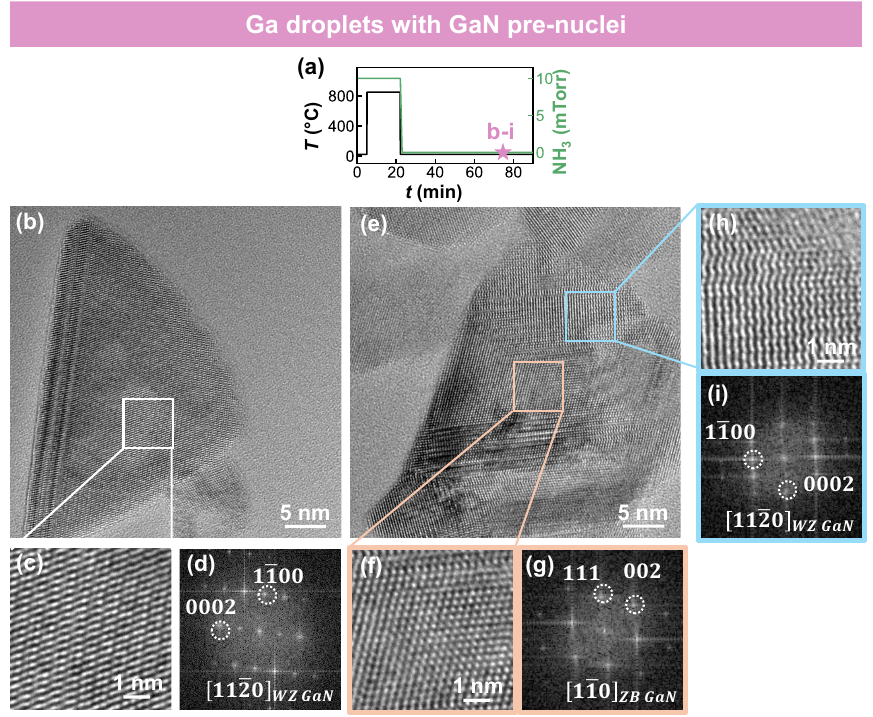}
    \caption{\label{fig:S2_ETEM} Nucleation and growth at high temperature: \textit{supplementary} in-situ TEM images during NH$_3$ exposure with $\Delta T / \Delta t = 400~^{\circ}\mathrm{C}/\mathrm{sec}$ and $T_{\text{target}}=850~\degree\text{C}$ (a) Heating and NH$_3$ exposure profiles for arrays of Ga droplets with GaN pre-nuclei. (b, e) HRTEM images of GaN QDs after NH$_3$ exposure. (c) provides a close-up view obtained from the white box in (b), while (f) and (h) provide close-up views obtained from the orange and blue boxes in (e), respectively. The FFTs taken from (c), (f), and (h) are indexed to (d) WZ GaN $[11\bar{2}0]$, (g) ZB GaN $[1\bar{1}0]$, and (i) WZ GaN $[11\bar{2}0]$, respectively.}
\end{figure}

\clearpage


\bibliography{References}